\documentclass[runningheads]{llncs}
\usepackage{graphicx}

\usepackage{amsmath}
\usepackage{pgfplotstable}

\usepackage[utf8]{inputenc} 
\usepackage{authblk}

\newcommand{\fref}[1]{Figure \ref{#1}}

\usepackage{amsfonts}

\usepackage{graphicx}
\graphicspath{{figures/}}
\DeclareGraphicsExtensions{.pdf,.jpeg,.png,.jpg}

\usepackage{hyperref} 
\usepackage[misc]{ifsym} 

\usepackage{listings}
\usepackage{xcolor}
\newcommand{\sslash}{\mathbin{/\mkern-6mu/}}

\usepackage{verbatim} 

\begin{document}
\raggedbottom

\title{Efficient Distributed Transposition \\Of Large-Scale Multigraphs And \\ High-Cardinality Sparse Matrices}

\titlerunning{Distributed Transposition of Multigraphs and High-Cardinality Matrices}

\author{
Bruno R. C. Magalh\~aes \inst{1} 
\and Felix Sch\"urmann\inst{1}\Letter  
}
\institute{
Blue Brain Project, Brain Mind Institute,\\\'Ecole polytechnique f\'ed\'erale de Lausanne (EPFL) \\ Campus Biotech, 1202 Geneva, Switzerland \\ \email{felix.schuermann@epfl.ch}
}


\maketitle

\begin{abstract}

Graph-based representations underlie a wide range of scientific problems. Graph connectivity is typically represented as a sparse matrix in the Compressed Sparse Row format. Large-scale graphs rely on distributed storage, allocating distinct subsets of rows to compute nodes.
Efficient matrix transpose is an operation of high importance, providing the reverse graph pathways and a column-ordered matrix view. 
This operation is well studied for simple graph models. 
Nevertheless, its resolution for multigraphs and higher-cardinality connectivity matrices is unexistent.

We advance state-of-the-art distributed transposition methods by providing a theoretical model, algorithmic details, MPI-based implementation and proof of mathematical soundness for such complex models. 
Benchmark results demonstrate ideal and almost ideal scaling properties for perfectly- and heterogeneously-balanced datasets, respectively.
\end{abstract}


\section{Introduction}

The Seven Bridges of K\"onigsberg, published by Leonhard Euler in 1736, is regarded as the first graph theory paper in history  \cite{gribkovskaia2007bridges}. The problem was to devise a walk across the city --- composed by two large islands connected to each other or to two mainland portions of the city by seven bridges ---  that would cross each bridge once and only once. Combined with Euler's formula relating the number of edges, vertices, and faces of a convex polyhedron, it represents the beginning of the mathematical branch known as topology \cite{wilson1985euler}. Since then, applications of graph theory in real life problems have grown extensively in most scientific domains. To name a few, computer science (pattern mining \cite{yan2002gspan}, image segmentation \cite{felzenszwalb2004efficient}), machine learning (sentiment analysis \cite{pang2004sentimental})
and biology (protein folding \cite{mohring1990graph})
. It is not uncommon for certain use cases to be represented by graph models in the order of billions of edges --- such as the travelling salesman problem on a large road map \cite{cornuejols1983halin}, friendship connectivity on social networks \cite{debnath2008feature} or URLs' cross-referral on web crawlers \cite{broder2000graph} --- requiring distributed storage on a network of compute nodes.

\begin{figure*}[t]
\centering
\includegraphics[width=0.95\textwidth]{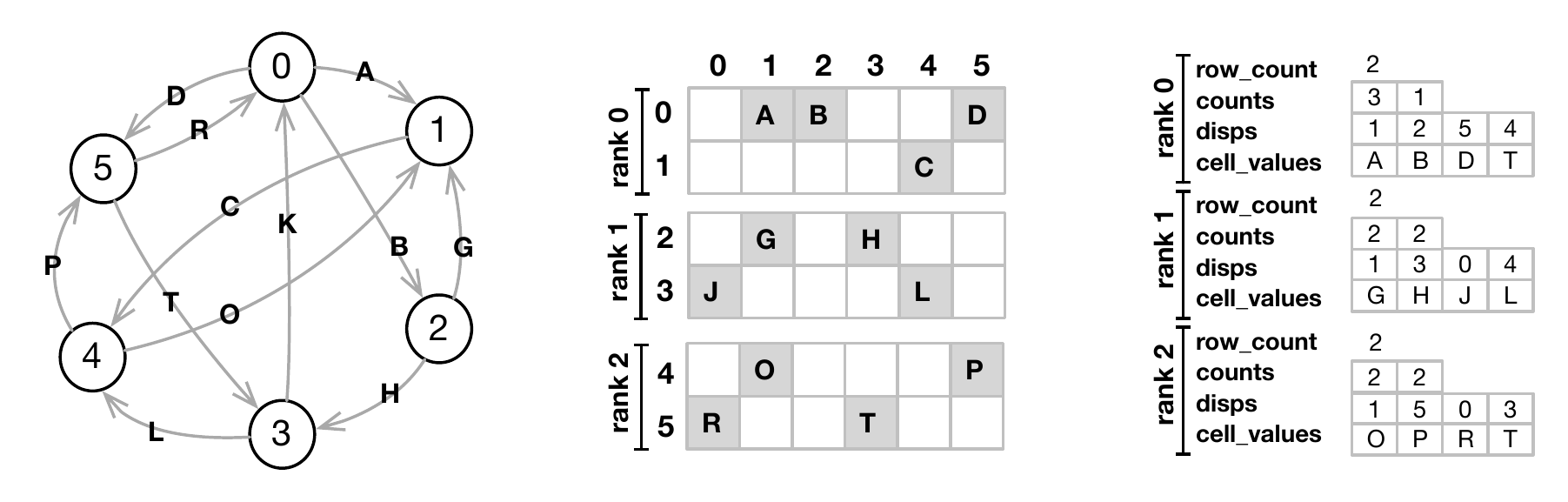}
\vspace{-0.2cm}
\caption{The Compressed Sparse Row format (CSR). \textbf{Left:} a sample directed graph with 6 vertices (nodes 0-5) with a single connection (edges A-Q) between nodes. \textbf{Middle:} the representative distributed sparse matrix, stored on three compute nodes (ranks 0-2). \textbf{Right:} The representative CSR format data arrays per compute node.
}
\label{pic-csr}
\end{figure*}

Graph connectivity is commonly represented as a matrix. Cells in the matrix represent edge information between two nodes uniquely identified by the row and column ids. For efficient storage, data is represented by the Compressed Sparse Row format (CSR, CRS or Yale format). The CSR format represents a sparse matrix by three arrays, that include the values of non-empty cells (or equivalently the non-missing nodes in the graph), the number of columns per row (number of connections for a given edge), and the indices of the columns (ids of connecting edges) --- refer to \fref{pic-csr} for an example. On a distributed memory environment, sequential rows are assigned to different compute nodes, and all columns within the row interval are stored in the same local machine. 
Graph partitioning \cite{karypis1995metis} allows for groups of non sequential rows to be assigned to compute nodes, minimising a given cost function. 

\begin{figure*}[t]
\centering
\includegraphics[width=1.0\textwidth]{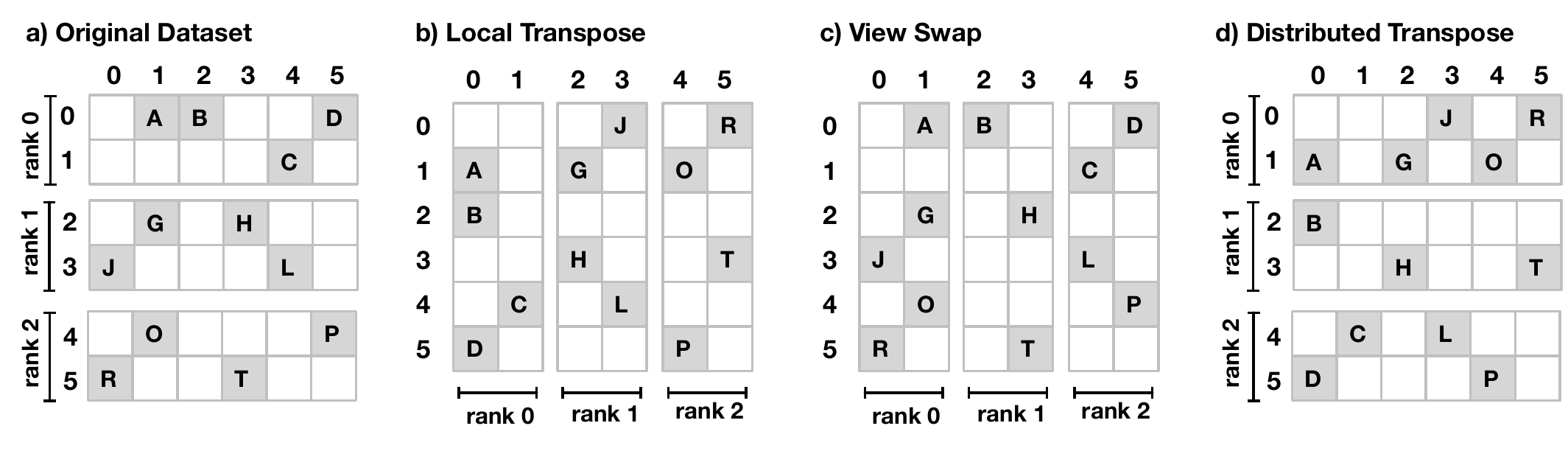}
\vspace{-0.5cm}
\caption{
Four distributed memory layouts for a sparse matrix with 12 cells placed on 6 rows equally distributed across 3 ranks. \textbf{a)} The original dataset stores a whole-row vertically-concatenated representation of the dataset. \textbf{b)} A locally transposed matrix holds the original dataset, represented by the orthogonal (horizontal) concatenation of rank datasets.  \textbf{c)} A view swap stores the same information as the original dataset in the orthogonal (whole-column) representation. \textbf{d)} The distributed transpose displays a transposed similar-view layout of the original dataset.
}
\label{pic-layouts}
\end{figure*}

The transpose of a matrix is an operation of high importance, as it provides (1) the reverse pathways between nodes of a graph, and (2) the retrieval in local memory of the alternative column-ordered connectivity, useful for iterations over rows of a given column, and vice-versa. The transpose operation is a solved problem for graph models with a single value per cell, or analogously, a single edge per pair of vertices. The Message Passing Interface (MPI) communication library, commonly used in large compute systems, exposes these operations via the \texttt{MPI\_Alltoall}, \texttt{MPI\_Alltoallv} and \texttt{MPI\_Alltoallw} calls on MPI-2, for the dense, homogeneously typed sparse, and heterogeneously typed sparse matrix use cases, respectively. Their asynchronous counterparts are provided by MPI-3 as \texttt{MPI\_Ialltoall}, \texttt{MPI\_Ialltoallv} and \texttt{MPI\_Ialltoallw}. The methods are based on all-to-all collective calls that performs a scatter-gather operation of the cells on a distributed CSR matrix. 
Alternative implementations focused on efficiency have been proposed, with an extensive analysis well covered by existing literature. To name a few, compressed sparse block transpose method from Buluc et al. \cite{bulucc2009parallel}, the methods for point-to-point communication and overlap of computation and communication from Choi et al. \cite{choi1995parallel}, the scan-based and transpose-merge-based methods from Wang et al. \cite{wang2016parallel}, the sorting-based methods form Gustavson et el. \cite{gustavson1978two} and the cache-efficient transpose methods based on the SIMD working pattern from Gustavson et al. \cite{gustavson2012parallel}, 
mostly suitable for smaller networks or networks with specialized point-to-point communication protocols, such as Infiniband.

Nevertheless, single edges between pairs of nodes is a property that is often unexistent  in real life situations. As an example, the matrix representing the list of common interests among pairs of users in a social network connectivity matrix; or the map of URLs cross-referral on a web crawler, where one may be required to store not only the number of hyperlinks between pages, but also a set of fields holding information about every link. For such scenarios, both a definition of a standard data format and the definition of the transpose operation or converse data matrix view have not yet been explored. 

 In that line of thought, we introduce a new class of matrix and graph transpose problems, represented by matrices with cell datatypes defined by list of values of distinct lengths (henceforth referred to as \textbf{high-cardinality matrices}), or equivalently graphs with multiple connections between node pairs, also referred to as \textbf{multigraphs}.  We describe a general data format for the distributed storage of the dataset, and the workflow of the distributed algorithm for the transpose of matrix and graph structures. We introduce the eXtended Compressed Sparse Row format (\textbf{XCSR}), an extension to the CSR format, and a serialized (array-based) representation of a dataset fitting the specification of our problem. Alongside the algorithmic description and proof of mathematical soundness, we provide an overview of existing methods, their implementation on the \texttt{C} programming language, and the contribution of our work to the state of the art. 
Our methods rely exclusively on MPI collective communication primitives, maximizing the efficiency in existing supercomputers with hardware specialized MPI implementation, taking advantage of network topology and retailer specific optimizations --- such as 6D torus, network backbones or Infiniband --- making our methods easily portable and efficient on a wide range of architectures. Benchmark results on a network of 128 Intel Xeon compute nodes demonstrate almost ideal weak and strong communication scaling properties on heterogeneously-balanced dataset, and ideal scaling on a perfectly-balanced scenario. 

\section{Problem Specification}
\label{sec-two}

A graph $G = (V, E)$ is a data structure consisting of a set of \textbf{nodes} or \textbf{vertices} $V$, and a set of links or \textbf{edges} $E$. An edge $\{i,j,e\} \subset \{\mathbb{N}, \mathbb{N}, T\}$ in $E$ represents a connection between vertices $i$ and $j$ in $V$ with edge information of type $T$, and can be either ordered or unordered, depending on whether the graph is directed on undirected.
A matrix $M$ represents the connectivity $E$ in $G$, iff for all $\{i,j,e\} \in E$, if $G$ is ordered then $M_{ij}=e$, and if is unordered then $M_{ij} = M_{ji} = e$. Since $M$ holds the possible connectivity between nodes in $V$, then $M$ is a square matrix of size $|V| \times |V|$. Moreover, if $M$ is unordered, then only a lower- or upper-diagonal representation is necessary, as the matrix is symmetric. For brevity, he following analysis focuses on the ordered use case only, as the unordered use case is simply a sub-problem of the ordered one.

A transposition of a matrix $M$ is defined by $M^T_{ij} = M_{ji}$, and holds an converse column-row cell placement of the initial matrix. The graph with connectivity described by a matrix $M^T$ can be then described simply by $G^{\star} = (V, E^{\star})$, where $E^{\star} = \big\{ \{j,i,e\}$ for all $\{i,j,e\} \in E \big\}$.

A distributed memory data layout assumes that the vertices $V$ are distributed across $R$ compute nodes (\textbf{ranks}) and only rows local to each memory region are directly accessible to a rank. We will refer to $G_{r} = (V_{r}, E_{r})$ as the subset of $G$ that is stored in rank $r$, with vertices $V_{r}$ and edges $E_{r}$. Each rank holds a disjoint subset of rows of the initial graph $G$, such that cover ( $\bigcup\limits_{r} G_{r} = G$) and distinct ($G_{r} \bigcap G_{s} = \emptyset, \forall r \neq s$) properties hold. Ranks only hold information about outgoing connectivity (or from edges in this rank to other vertices), i.e. $E_{r} = \Big\{ \{i,j,e\} \in E \text{ } \Big| \text{ } i \in V_r \Big\}$. Thus, the same cover and disjoint properties also hold for edges.
  
Given a dense matrix representation of a graph connectivity $E$, the algorithm to compute $G^{\star}_{r} = (V_{r}, E^{\star}_{r})$ for every rank $r$, requires only the computation of the matrix $M^{T}_{r}$ for the connectivity $E^{\star}_{r}$. Note that vertices information is local to a compute node, 
 therefore the nodes information $V$ is the same for $G$ and $G^{\star}$. The implementation of the distributed transpose is well known for a \textbf{dense} connectivity matrix, and available via the \texttt{MPI\_Alltoall} collective call, that inputs (outputs) an array of values, the size of the array, and datatype of the values to be sent (received), with an additional term for the network communicator that is an abstraction of the ranks in the network:
 \\
 \begin{minipage}{\textwidth}
\begin{lstlisting}
int MPI_Alltoall(                                         //output status
    void *sendbuf, int sendcount, MPI_Datatype sendtype,  //input values
    void *recvbuf, int recvcount, MPI_Datatype recvtype,  //output values
    MPI_Comm comm);                                       //input comm
\end{lstlisting}
\end{minipage}

It is convenient to provide the typing of input and output parameters, so that the explanation of the methods that follow are defined more clearly. The typing of \texttt{MPI\_Alltoall} is defined as:
\vspace{0.2cm}\\{\small $MPI\_Alltoall: ( T^N \times \mathbb{N} \times T_{MPI} ) \times C_{MPI} \rightarrow ( T^N \times \mathbb{N} \times T_{MPI}) \times S_{MPI}$}\\\vspace{-0.3cm}\\
where $T^N$ represents an array of values of type $T$ generalized as a \texttt{void} pointer, $T_{MPI}$ is the MPI-defined datatype related to $T$ in the \texttt{MPI\_Datatype} collection, $C_{MPI}$ is the user or MPI-defined identifier for the communicator \texttt{MPI\_Comm} (typically \texttt{MPI\_COMM\_WORLD} for all ranks), and $S_{MPI}$ is the error status described by the \texttt{int} return value.
Note that the value for the \texttt{sendcount} and \texttt{recvcount} variables are equal to the network size $R$ --- easily retrievable via \texttt{MPI\_Comm\_size} --- and the types \texttt{sendtype} and \texttt{recvtype} are the same for the transpose. 
The asynchronous variant \texttt{MPI\_Ialltoall} includes an extra term \texttt{MPI\_Request *request} for probing of status and will be omitted as the logic follows analogously.

The transpose of the \textbf{sparse} matrix counterpart is a solved problem. The algorithm is as follows: 
\begin{enumerate}
\item Each rank is required  to hold the init and end row index of every rank, so that it can match column id with target rank when transposing. This can be performed at the onset of execution by a collective gathering (\texttt{MPI\_Allgather}) of the number of rows per rank, followed by computation of offsets, where each rank's offset is the sum of the previous ranks' row count;
\item The rank offsets allow for a rank $r$ to compute the amount of values to be sent to each rank (including itself). This all-to-all count is represented by a dense distributed matrix of size $R\times R$. A transpose of the counts matrix using the previous \texttt{MPI\_Alltoall} method yields a local representation of the amount of data to be sent and received by all ranks;
\item a scatter-gather operation follows and sends (delivers) of cell values to the correct recipients, based on the previous sent (received) counts. A rank $r$ is now able to retrieve the layout of the locally transposed matrix $M^T_r$ by mapping the received values to the row offset of each rank;
\end{enumerate}

The final collective all-to-all variable size communication method is implemented by MPI in the \texttt{MPI\_Alltoallv} function:
\\
\begin{minipage}{\textwidth}
\begin{lstlisting}
int MPI_Alltoallv(
    void *sendbuf, int *sendcounts, int *sdispls, MPI_Datatype sendtype,
    void *recvbuf, int *recvcounts, int *rdispls, MPI_Datatype recvtype,
    MPI_Comm comm);
\end{lstlisting}
\end{minipage}
with the typing:
\vspace{0.2cm}\\{\small $MPI\_Alltoallv: ( T^N \times \mathbb{N}^M \times \mathbb{N}^M \times T_{MPI}) \times C_{MPI}  \rightarrow (T^N \times \mathbb{N}^M \times \mathbb{N}^M  \times T_{MPI}) \times S_{MPI}$}.

Two variations of the previous method are possible and of direct implementation: (1) for user-defined homogeneous datatypes on edges, one can either create an MPI derived data struct or provide a serialization of an object --- with the $T_{MPI}$ datatype set to \texttt{MPI\_BYTE} and the counts and offsets scaled linearly to the byte size of datatype; (2) for heterogeneous datatypes across edges, a generalized version is possible with \texttt{MPI\_Alltoallw}, providing the list of ltypes $T^\mathbb{N}_{MPI}$ in \texttt{MPI\_Datatype *sendtypes} and \texttt{*recvtypes}, as in:
\\
 \begin{minipage}{\textwidth}
\begin{lstlisting}
int MPI_Alltoallw(
    void *sendbuf, int *sendcounts, int *sdispls, MPI_Datatype *sendtypes,
    void *recvbuf, int *recvcounts, int *rdispls, MPI_Datatype *recvtypes,
    MPI_Comm comm);
\end{lstlisting}
\end{minipage}
with the parameter typing adjusted accordingly:
\vspace{0.2cm}\\{\small $MPI\_Alltoallw: ( T^N\times\mathbb{N}^M \times \mathbb{N}^M \times T_{MPI}) \times C_{MPI}  \rightarrow (T^N \times \mathbb{N}^M \times \mathbb{N}^M  \times T_{MPI}) \times S_{MPI}$}\vspace{-0.3cm}\\

The description that follows focuses on the \texttt{MPI\_Alltoallv} use case, as the implementation of the variations from this base case is straightforward. A transpose of a sparse distributed matrix is a sequence of three collective communication calls: an \texttt{MPI\_Allgather} for the collection of row counts and posterior computation of rank offsets, an \texttt{MPI\_Alltoall} dense matrix transpose for the communication of value counts to be sent and received, and an \texttt{MPI\_Alltoallv} for the transpose of the sparse matrix transpose that exchanges the cell values. The transpose method can thus be encapsulated in a single function call as:
\\
 \begin{minipage}{\textwidth}
\begin{lstlisting}
int Transpose( int row_count,
    void *sendbuf, int *sendcounts, int *sdispls, MPI_Datatype sendtype,
    void *recvbuf, int *recvcounts, int *rdispls, MPI_Datatype recvtype,
    MPI_Comm comm);
\end{lstlisting}
\end{minipage}
with the typing:
{\vspace{0.2cm}\\{\small $Transpose: \mathbb{N} \times (T^N \times \mathbb{N}^M \times \mathbb{N}^M \times T^N_{MPI}) \times C_{MPI} \rightarrow ( T^N   \times \mathbb{N}^M \times \mathbb{N}^M \times T^N_{MPI}) \times S_{MPI}$}\vspace{-0.3cm}\\


\begin{figure*}[t]
\centering
\includegraphics[width=1.0\textwidth]{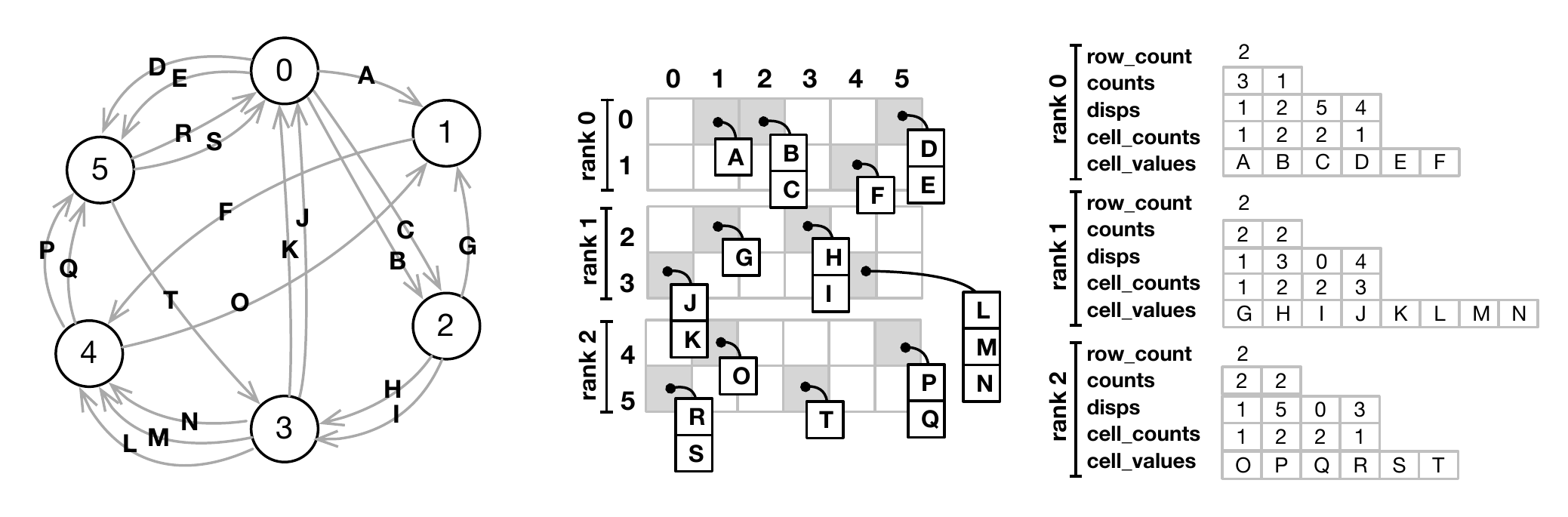}
\vspace{-0.6cm}
\caption{The extended Compressed Sparse Row format (XCSR). \textbf{Left:} a sample graph with 6 vertices (0-5) and multiple edges (A-Q) between nodes. \textbf{Middle:} the representative distributed sparse matrix, stored in 3 ranks. \textbf{Right:} The XCSR data arrays.
}
\label{pic-xcsr}
\end{figure*}

For readability purposes, we simplify the previous notation by: (1) enforcing similar send and receive datatypes; (2) inserting a referenced (\texttt{\&}) declaration of rend/recv buffers for an \textit{in-place} alteration of arguments, as the pre-transpose information is not required after the transpose; and (3) performing a strongly-typed declaration of \texttt{void*} buffers into a user-specified type \texttt{T}, leading to:
\\
 \begin{minipage}{\textwidth}
\begin{lstlisting}
template<typename T>
int Transpose(                                       //output status
    int row_count,                                   //input row count
    T *& cell_values, int *& counts, int *& displs,  //input/output matrix
    MPI_Comm comm);                                  //input comm
\end{lstlisting}
\end{minipage}
with data typing simplified to:
{\vspace{0.2cm}\\{\small $Transpose: \mathbb{N} \times (T^N \times \mathbb{N}^M \times \mathbb{N}^M) \times C_{MPI} \rightarrow ( T^N   \times \mathbb{N}^M \times \mathbb{N}^M) \times S_{MPI}$}\vspace{-0.3cm}\\

The aforementioned header definition provides a one-to-one match to the Compressed Sparse Row format. For completion, refer to \fref{pic-csr} for a sample graph, its representative sparse matrix and the data structures local to each rank. Existing transpose methods cover extensively the transpose operation for such matrix representation, either at the CSR format, the Block Compressed Format \cite{bulucc2009parallel}, or other equivalent representation. We now enter our problem domain by extending the formalism to higher-cardinality cells (i.e. matrix cells whose cells hold several values), and multigraphs' connectivity matrices.

Suppose nodes connectivity is not defined by $\{i,j,e\}  \subset \{\mathbb{N}, \mathbb{N}, T\}$ for an edge of type $T$ as before, but instead by $\{\mathbb{N}, \mathbb{N}, T^{N_{ij}}\}$, where edge information is defined by a list of connections of \textbf{variable length} $N_{ij}$. To comply with the new cell specification, we represent the serialized cell values by the array of all values and the count of values per cell, typed $T^{\mathbb{N}} \times \mathbb{N}^\mathbb{N}$, thus adding an extra \texttt{cell\_counts} argument to the previous header definition:
\\
\begin{minipage}{\textwidth}
\begin{lstlisting}
template<typename T>
int Transpose(
    int row_count,
    T *& cell_values, int *& counts, int *& displs, int *& cell_counts,
    MPI_Comm comm);
\end{lstlisting}
\end{minipage}
with parameters typing following accordingly as:
{\vspace{0.2cm}\\{\small $Transpose: \mathbb{N} \times (T^{\mathbb{N}} \times \mathbb{N}^\mathbb{N} \times \mathbb{N}^\mathbb{N} \times \mathbb{N}^\mathbb{N}) \times C_{MPI} \rightarrow ( T^{\mathbb{N}}   \times \mathbb{N}^\mathbb{N} \times \mathbb{N}^\mathbb{N} \times \mathbb{N}^\mathbb{N}) \times S_{MPI}$}.\vspace{-0.3cm}\\

We refer to the CSR format with the additional term as the eXtended Compressed Sparse Row format (\textbf{XCSR}, illustrated in \fref{pic-xcsr}). Analogously to CSR-based transpositions, the aforementioned transpose method is XCSR-compatible and of general application to any problem described by a serializeable cell defined by list of values of type $T$.

The C++ implementation of the \texttt{Transpose} method for the XCSR data structure is available at the Blue Brain open source repository \cite{TransposerGitHub}. For brevity, will be omitted from this document. Instead, the description of the algorithm and the proof of mathematical soundness are provided in the following section.

\section{Algorithm}

We start the formulation of our problem resolution with the mathematical formalism underlying the distributed matrix transpose operations. A \textbf{horizontal concatenation} of two matrices $M_{n \times m}$ and $N_{n \times m'}$ is represented by $M \| N$ and defined as the operation to join two sub-matrices horizontally into a matrix of dimensionality ${n \times (m+m')}$, such that: 
\begin{equation}
(M \| N)_{i j} =\left\{
  \begin{array}{@{}ll@{}}
    M_{i \text{ } j} & \text{, if } j \leq m\\
    N_{i \text{ } j-m} & \text{, otherwise}
  \end{array}\right.
\end{equation}

Analogously, \textbf{a vertical concatenation} $M_{n \times m} \sslash N_{n' \times m}$ joins vertically two sub-matrices into a matrix of dimensionality ${(n+n') \times m}$, such that: 
\begin{equation}
(M \sslash N)_{i j} =\left\{
  \begin{array}{@{}ll@{}}
    M_{i \text{ } j} & \text{, if } i \leq n\\
    N_{i-n \text{ } j} & \text{, otherwise}
  \end{array}\right.
\end{equation}

\begin{figure*}[t]
\centering
\includegraphics[width=1.0\textwidth]{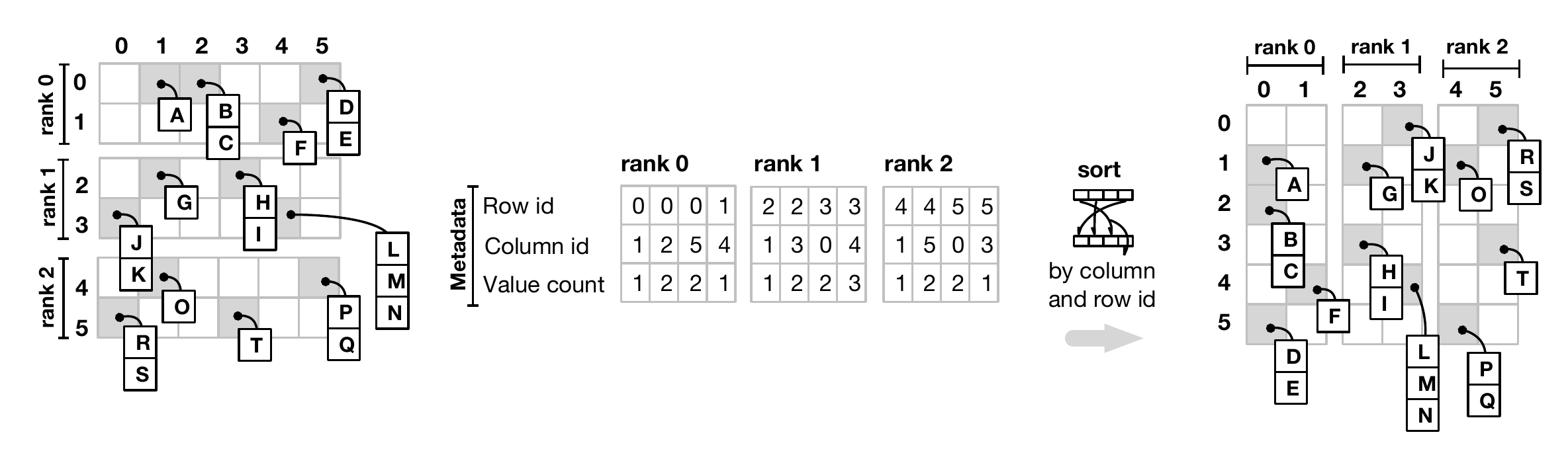}
\vspace{-0.6cm}
\caption{\textbf{Left:} a distributed matrix, illustrative of the data structures in \fref{pic-xcsr}; \textbf{Centre:} the corresponding metadata, storing row id, column id and number of values per cell; \textbf{Right:} the locally transposed representation of the matrix.
}
\label{pic-matrix-transposition-1}
\end{figure*}

We refer to a \textbf{view} as the perspective of data storage: the column view describes the matrix as the vertical concatenation of the subsets of rows stored on each rank. The row view represents the horizontal concatenation of subsets of columns on each rank. It follows that $(M \| N)^T = (N^T \sslash M^T)$ and $(M \sslash N)^T = (N^T \| M^T)$, as both concatenations provide the same dataset yet described by two alternative views, and 
the transpose of a concatenated dataset yields the orthogonally-concatenated dataset. 
The \textbf{local transpose} of a matrix $M$ represented by the concatenation of $R$ submatrices in either view, is defined by the concatenation of the transpose of the individual matrices in the orthogonal view:
\begin{equation}
LocalTranspose(M)_{i j} =\left\{
  \begin{array}{@{}ll@{}}
    \big( M^T_1 \| M^T_2 \| ... \| M^T_R \big)_{j i}& \text{, if $M$ is in row view }\\
    \big( M^T_1 \sslash M^T_2 \sslash ... \sslash M^T_R \big)_{j i}& \text{, otherwise}
  \end{array}\right.
\end{equation}

A sample application of the local transpose function is displayed in \fref{pic-layouts}, layout b). The algorithm that describes the local transpose of the XCSR to a dataset is detailed in \fref{pic-matrix-transposition-1}. It is relevant to mention that the operation yields a transposed version of the original rank matrices that formed the initial dataset, thus no communication between ranks is necessary as the data $M_r$ of every rank is locally transposed into $M^T_r$. Moreover, it is an involutory function as both views are orthogonal thus $(LocalTranspose \cdot LocalTranspose) (M) = M$. 

The \textbf{view swap} of a distributed matrix that alternates between a data representation on a view and its orthogonal (from row- to column-accessible and vice-versa), while maintaining the same matrix contents is defined by:
\begin{equation}
ViewSwap(M)_{i j} =\left\{
  \begin{array}{@{}ll@{}}
    \big( M_1 \| M_2 \| ... \| M_R \big)_{i j}& \text{, if $M$ is in row view}\\
    \big( M_1 \sslash  M_2 \sslash  ... \sslash  M_R \big)_{i j}& \text{, otherwise}
  \end{array}\right.
\end{equation}

\begin{figure*}[t]
\centering
\includegraphics[width=1.0\textwidth]{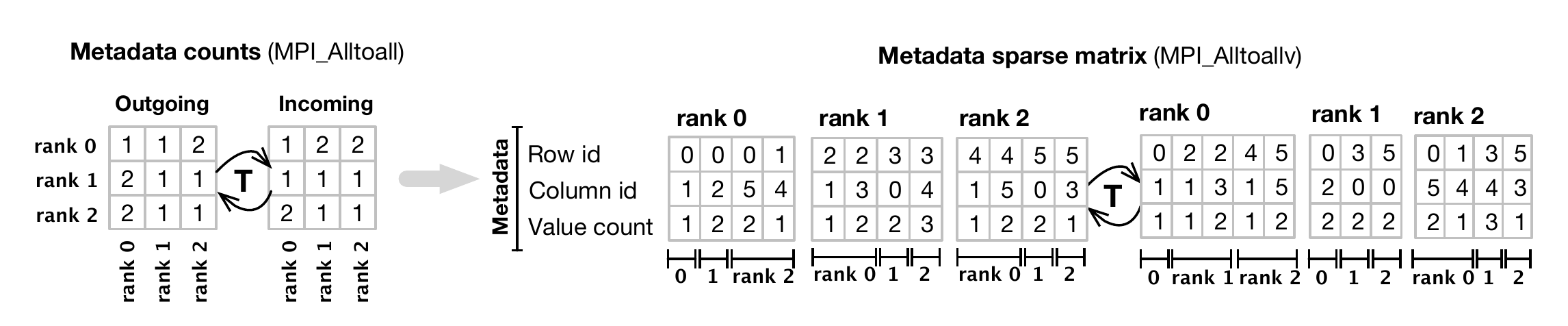}
\vspace{-0.5cm}
\caption{View swap, step 1/2, continuing \fref{pic-matrix-transposition-1}: the communication of the metadata. \textbf{Left:} the transposition of the dense matrix holding metadata counts to be sent/recv per rank. \textbf{Right:} The communication of the metadata via a sparse matrix transpose.
}
\label{pic-matrix-transposition-2}
\end{figure*}

An application of the view swap is illustrated by layout c) in \fref{pic-layouts}. At the level of a rank, the matrix data layout after a view swap is unknown until the swap is performed, as ranks do not hold information about the matrix structure across other ranks. Therefore, as mentioned previously, a communication step is required to be executed beforehand, in order to gather the number of rows held by individual ranks, and compute the row intervals on every other rank $r$ as $\big[\sum^{r-1}_{r=1} |V_r|, \text{ } \sum^{r}_{r=1} |V_r| \big)$. This information is required for the correct matching of column/row id to target rank, used in the sparse transposition steps that follow.

The view swap algorithm follows then in two communication steps. The first step exchanges the local structure of the matrix, with a dense and distributed sparse matrix transpose operation communicating the count and the metadata structures, respectively --- refer to \fref{pic-matrix-transposition-2} for further details. As a side note, sparse matrix transpose of regular matrices would be finalized at this step, as it provides enough information for the use case of single value per cell (by defining the tuple $<$row id, column id, cell value$>$ as metadata in Fig. \ref{pic-matrix-transposition-2}). Moreover, for the use cases of identical sizes across cells, transposition could be completed by serializing cell values as a unique data structure, similarly to existing sparse transpose methods.  For our use case of XCSR with higher cardinality per cell and distinct cell lengths, an additional step is required.

\begin{figure*}[t]
\centering
\includegraphics[width=1.0\textwidth]{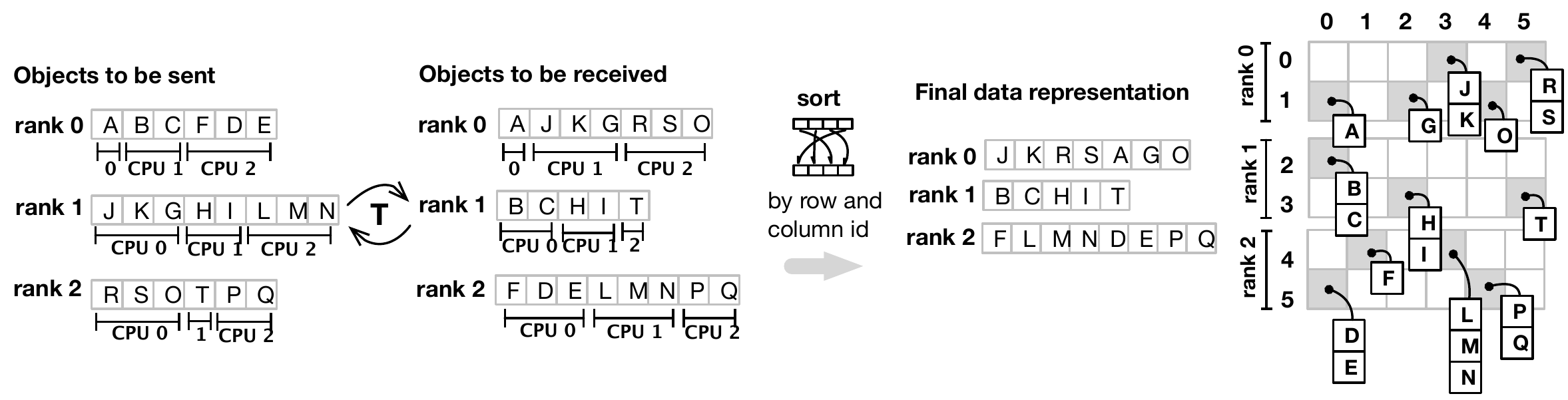}
\vspace{-0.5cm}
\caption{View swap, step 2/2, continuing \fref{pic-matrix-transposition-2}: the communication of cell values. \textbf{Left:} A dense transpose of value counts per recipient, followed by a sparse transpose of values, delivers the matrix content to the correct recipient ranks. \textbf{Right:} A row-column ordering of the received data based on the metadata exchanged previously (Fig. \ref{pic-matrix-transposition-2}) yields the distributed transpose of the original matrix.
}
\label{pic-matrix-transposition-3}
\end{figure*}

The second step of the view swap performs an all-to-all communication of value counts to be sent/received, followed by a selective scatter-gather of the values. A local reordering of the data structure follows, using a row-column order of values, according to the matrix structure exchanged by the transpose in the previous step. This process is detailed in \fref{pic-matrix-transposition-3}. As a relevant remark, the view swap method is also an involutory function, as each view swap performs two involutory transpose operations, and two consecutive view swaps yield the initial dataset.

The \textbf{distributed sparse transpose} (layout d in \fref{pic-layouts}) is defined by a composition of the local transpose and view swap methods. We have shown that both the local transpose and the view swap are XCSR compatible.  We will show the distributed sparse transpose is also mathematically sound when applied to the XCSR. Take a matrix $M$ in row view, represented by the vertical concatenation of partial matrices $M_1 \sslash M_2 \sslash ... \sslash M_R$.  A local transpose of $M$ leads to $M^T_{i j} = ( M^T_1 \| M^T_2 \| ... \| M^T_R)_{ji}$. Applying a view swap, we have $M^T_{i j} = ( M^T_1 \sslash M^T_2 \sslash ... \sslash N^T_R)_{ji}$, which is the definition of the distributed transpose of $M$ in the original view. In brief, it follows that $Transpose (M) = (LocalTranspose \cdot ViewSwap) (M)$. The verification of the commuted involutory functions with $Transpose (M) = (ViewSwap \cdot LocalTranspose) (M)$ also holds: a view swap over the original dataset leads to  $M_{i j} = ( M_1 \sslash M_2 \sslash ... \sslash M_R)_{i j}$. The composition of a local transpose yields the final result $M^T_{i j} = ( M^T_1 \sslash M^T_2 \sslash ... \sslash N^T_R)_{ji}$. As $Transpose$ is a composition of two XCSR-compatible involutory functions, and because both functions commute i.e. $(ViewSwap \cdot LocalTranspose) = (LocalTranspose \cdot ViewSwap)$, then $Transpose$ is by definition XCSR-compatible and involutory \cite{kubrusly2011elements}. 

To finalize, the full transpose algorithm requires five collective communication calls, performing an \texttt{MPI\_Allgather} to compute the row offsets of ranks, two \texttt{MPI\_Alltoall} and two \texttt{MPI\_Alltoallv} for the metadata and cell values transpositions. Moreover, the general applicability of the XCSR and the involutory property of our transpose method, are two properties of high importance as they guarantee that distributed transposition can be validly executed any number of times, while respecting the data integrity of a XCSR-based graph problem representation. 

\section{Benchmark}

We benchmarked our methods on a network of 128 Intel Xeon Gold 6140 compute nodes with 18-core at 2.3GHz with AVX-512, interconnected by two Infiniband EDR networks with a communication bandwidth of 100Gbit/s. Communication methods are provided by the mvapich2 MPI implementation.


Our initial testbench measured the runtime of our methods applied to a highly imbalanced distributed matrix. 
The input data distribution is as follows: each row holds a total 300 thousand to 1 million columns, uniformly distributed; each matrix cell stores a list of 128-byte values, with a mean of 5 values per cell. To test the involutory properties of our methods, each execution performs a composition of 12 transpose operations on the same dataset. As a relevant remark, the computation involved in the transposition process is considered negligible as the runtime is dictated mostly by the communication workload. 
The efficiency of the benchmark in terms of weak and strong scaling on a logarithmic axis representation is presented in \fref{pic-weak-strong}. The weak scaling benchmark presents the variation of runtime with the number of ranks (4 to 128) for a fixed problem size per rank. The strong scaling presents the variation of runtime in the same network configuration, for a fixed total problem size. Results suggest that the execution time increases almost linearly with the input size (mean number of rows) per rank on the weak scaling analysis, and with total input size on the strong scaling use case. The rationale is straightforward, as the runtime is proportional to the bandwidth which is dictated by the number of rows per rank. Yet, the number of columns varies extensively across nodes, and consequently across ranks, leading to a heterogeneous amount of communication per rank on the collective calls. Weak and strong scaling properties follow very closely the ideal patterns of communication scaling, discussed next.

\begin{figure*}[t]
\centering
\includegraphics[width=1.\textwidth]{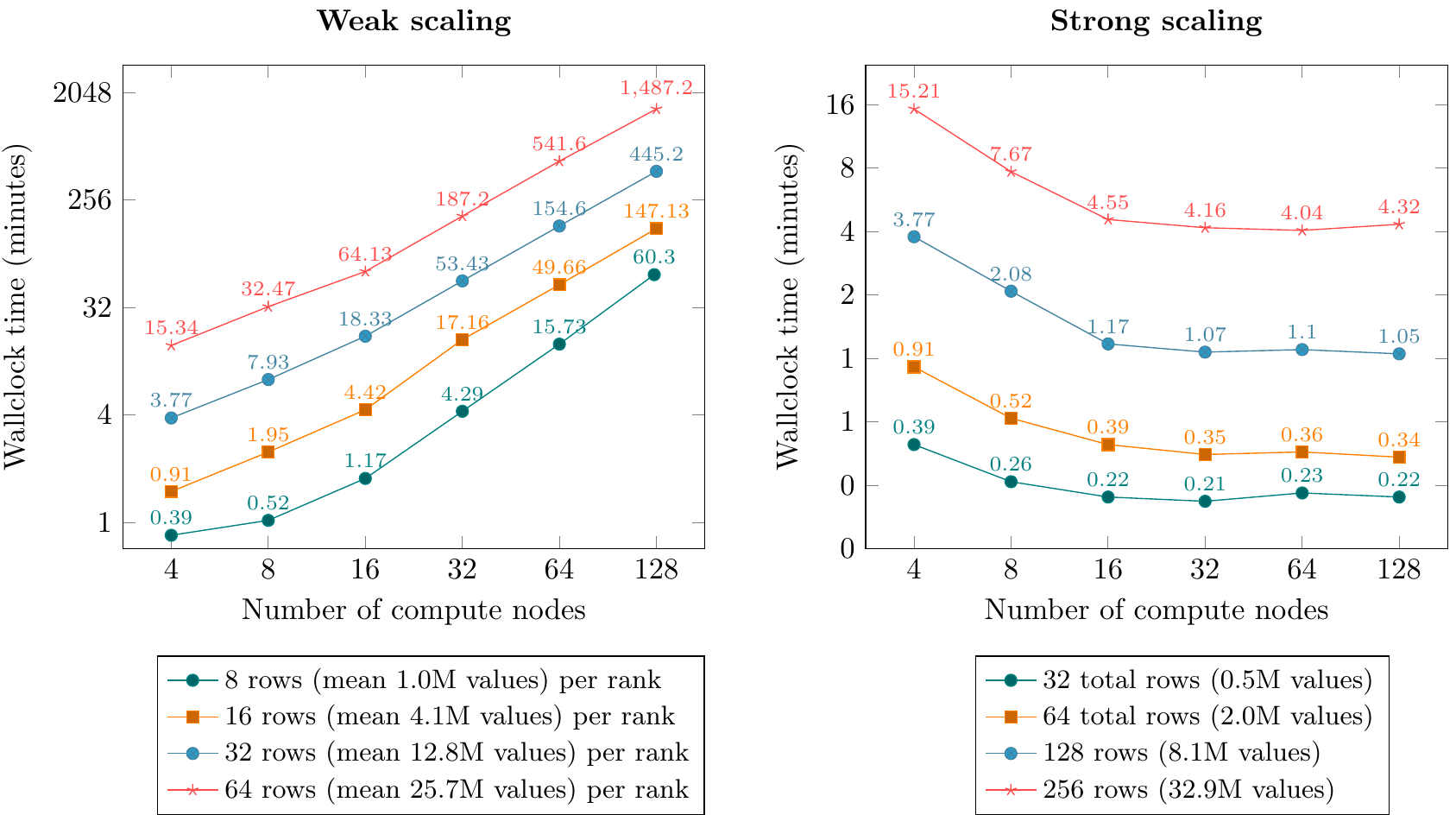}
\caption{Weak and strong scaling benchmarks of the transpose algorithm applied to a heterogeneously-balanced dataset.
}
\label{pic-weak-strong}
\end{figure*}

\begin{figure*}[t]
\centering
\includegraphics[width=.95\textwidth]{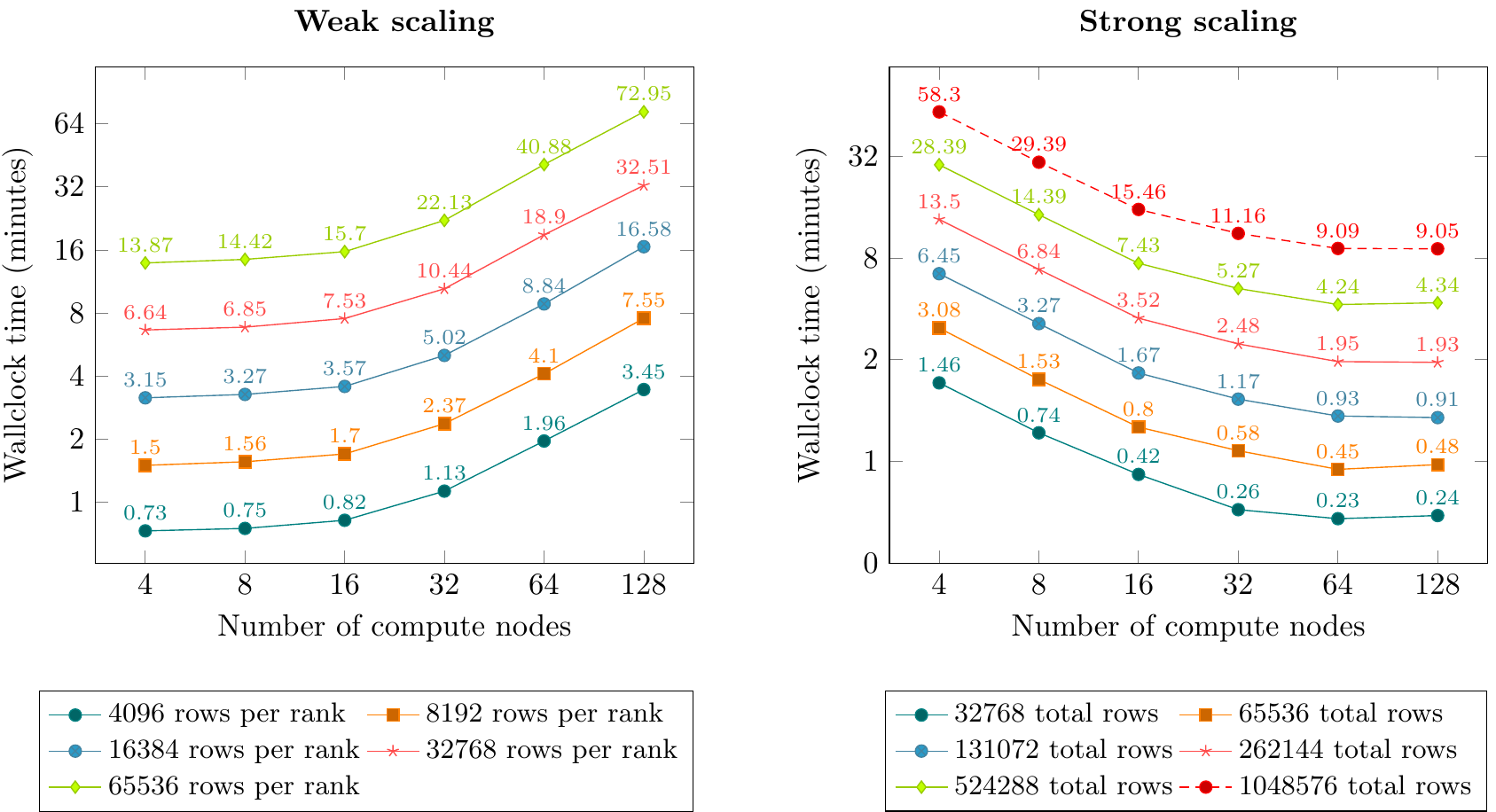}
\caption{Weak and strong scaling benchmarks for perfectly balanced communication across compute nodes. A row contains 512 cells, each represented by 10 integer values.
}
\label{pic-weak-strong-general}
\end{figure*}

An analysis of scaling properties in ideal conditions was performed on a similar benchmark applied a perfectly balanced dataset. The testbench transposes a matrix with cells represented by lists of 10 integer values. Each row contains 512 columns.  Each rank is assigned an increasing data set of 4096 to 65536 rows on the weak scaling use test, and an equivalent sum of rows for the several inputs profiled on the strong scale benchmarks. The data distribution guarantees balanced datasets and communication across ranks, and a linear increase of communication with the matrix size. Further details and the source code of the testbench is available alongside the implementation of the transposing algorithm \cite{TransposerGitHub}. The benchmark results are displayed in \fref{pic-weak-strong-general}.
We aim to show ideal weak and strong scaling properties of our algorithm. As the runtime is dictated mostly by collective communications, the scaling of the methods are heavily dependent on the efficiency of the underlying network and MPI implementation. Results suggest a linear increase of runtime with the network size for 32 nodes or more, on the weak scaling analysis, independently of the input size. Similarly, the strong scaling runtimes decrease to a pattern of constant execution time as the network size increases to 32 nodes or more.

These results agree with the ideal communication pattern for weak and strong scaling previously demonstrated by Hoefler et al. \cite{hoefler2010toward} --- presenting similar runtimes with linearly increasing weak scaling and constant strong scaling when plotted on a log-log axis representation --- demonstrated on a network of up to 32K compute nodes. It is relevant to emphasize that  ideal communication patterns do not follow the common assumption of ideal weak and strong scaling in computation, modelled by constant and linearly decreasing runtimes on the weak and strong scaling axes, respectively. The rationale behind this discrepancy is detailed in the aforementioned bibliographic reference: in brief, for a given input size per compute node, the computation time is reduced when increasing the network size due to added compute parallelism, yet the total data transmitted and communication involved in collective communication calls remains constant.

\section{Conclusion}
\label{sec-conclusion}

This paper introduced a new class of algorithms for the parallel transposition of distributed multigraphs and distributed sparse matrices with higher-cardinality cell values. We showed that the existing standard representations are not sufficient to hold high-cardinality information. To overcome this limitation, we introduced an extended version of the Compressed Sparse Row commonly utilised for graph connectivity representation. 
We detailed the distributed dense matrix, sparse matrix, and multigraph use cases implementation on the \texttt{C/C++} programming languages. For full coverage of the topic, we provided guidances on the implementation of our methods 
on sparse matrices or multigraphs with heterogeneously-typed cell and edge values. Methods provided were completed with their analytical formulation and proof of mathematical soundness.
Algorithms were formalized using first principles of computer science, and relying only on existing MPI collective communication calls, making it highly efficient and of direct porting to modern supercomputing architectures.

We performed a weak and strong scaling benchmark of our methods on a network of 128 Intel Xeon 6140 compute nodes connected by an Infiniband network interface.
Benchmark results on a perfectly-balanced dataset demonstrated linearly-increasing weak scaling, and constant strong scaling patterns on a logarithmic plot representation, suggesting ideal weak and strong communication patterns for large network sizes, independently of the input size. Runtimes of heterogeneously-balanced datasets 
displayed almost ideal scaling properties, demonstrating the feasibility of our methods on a wide range of scientific problem domains represented by large-scale graph or matrix data structures.

\section*{Acknowledgements}

This study was supported by funding to the Blue Brain Project, a research center of the École polytechnique fédérale de Lausanne (EPFL), from the Swiss government’s ETH Board of the Swiss Federal Institutes of Technology. The authors would like to thank Francesco Cremonesi for technical discussions. 

\newpage
\bibliographystyle{splncs04}
\bibliography{isc2014,hpc,functionalizer}

\begin{thebibliography}{10}
\providecommand{\url}[1]{\texttt{#1}}
\providecommand{\urlprefix}{URL }
\providecommand{\doi}[1]{https://doi.org/#1}

\bibitem{broder2000graph}
Broder, A., Kumar, R., Maghoul, F., Raghavan, P., Rajagopalan, S., Stata, R.,
  Tomkins, A., Wiener, J.: Graph structure in the web. Computer networks
  \textbf{33}(1-6),  309--320 (2000)

\bibitem{TransposerGitHub}
{Bruno Magalhaes and The Blue Brain Project}: Distributed transposer of sparse
  matrices with high-cardinality cell structures.
  \url{https://github.com/bluebrain/matrix-transposer} (2018)

\bibitem{bulucc2009parallel}
Bulu{\c{c}}, A., Fineman, J.T., Frigo, M., Gilbert, J.R., Leiserson, C.E.:
  Parallel sparse matrix-vector and matrix-transpose-vector multiplication
  using compressed sparse blocks. In: Proceedings of the twenty-first annual
  symposium on Parallelism in algorithms and architectures. pp. 233--244. ACM
  (2009)

\bibitem{choi1995parallel}
Choi, J., Dongarra, J.J., Walker, D.W.: Parallel matrix transpose algorithms on
  distributed memory concurrent computers. Parallel Computing  \textbf{21}(9),
  1387--1405 (1995)

\bibitem{cornuejols1983halin}
Cornu{\'e}jols, G., Naddef, D., Pulleyblank, W.R.: Halin graphs and the
  travelling salesman problem. Mathematical programming  \textbf{26}(3),
  287--294 (1983)

\bibitem{debnath2008feature}
Debnath, S., Ganguly, N., Mitra, P.: Feature weighting in content based
  recommendation system using social network analysis. In: Proceedings of the
  17th international conference on World Wide Web. pp. 1041--1042. ACM (2008)

\bibitem{felzenszwalb2004efficient}
Felzenszwalb, P.F., Huttenlocher, D.P.: Efficient graph-based image
  segmentation. International journal of computer vision  \textbf{59}(2),
  167--181 (2004)

\bibitem{gribkovskaia2007bridges}
Gribkovskaia, I., Halskau~Sr, {\O}., Laporte, G.: The bridges of
  k{\"o}nigsberg—a historical perspective. Networks: An International Journal
   \textbf{49}(3),  199--203 (2007)

\bibitem{gustavson2012parallel}
Gustavson, F., Karlsson, L., K{\aa}gstr{\"o}m, B.: Parallel and cache-efficient
  in-place matrix storage format conversion. ACM Transactions on Mathematical
  Software (TOMS)  \textbf{38}(3), ~17 (2012)

\bibitem{gustavson1978two}
Gustavson, F.G.: Two fast algorithms for sparse matrices: Multiplication and
  permuted transposition. ACM Transactions on Mathematical Software (TOMS)
  \textbf{4}(3),  250--269 (1978)

\bibitem{hoefler2010toward}
Hoefler, T., Gropp, W., Thakur, R., Tr{\"a}ff, J.L.: Toward performance models
  of mpi implementations for understanding application scaling issues. In:
  European MPI Users' Group Meeting. pp. 21--30. Springer (2010)

\bibitem{karypis1995metis}
Karypis, G., Kumar, V.: Metis--unstructured graph partitioning and sparse
  matrix ordering system, version 2.0  (1995)

\bibitem{kubrusly2011elements}
Kubrusly, C.S.: Elements of operator theory. Springer (2011)

\bibitem{mohring1990graph}
M{\"o}hring, R.H.: Graph problems related to gate matrix layout and pla
  folding. In: Computational graph theory, pp. 17--51. Springer (1990)

\bibitem{pang2004sentimental}
Pang, B., Lee, L.: A sentimental education: Sentiment analysis using
  subjectivity summarization based on minimum cuts. In: Proceedings of the 42nd
  annual meeting on Association for Computational Linguistics. p.~271.
  Association for Computational Linguistics (2004)

\bibitem{wang2016parallel}
Wang, H., Liu, W., Hou, K., Feng, W.c.: Parallel transposition of sparse data
  structures. In: Proceedings of the 2016 International Conference on
  Supercomputing. p.~33. ACM (2016)

\bibitem{wilson1985euler}
Wilson, P.R.: Euler formulas and geometric modeling. IEEE Computer Graphics and
  Applications  \textbf{5}(8),  24--36 (1985)

\bibitem{yan2002gspan}
Yan, X., Han, J.: gspan: Graph-based substructure pattern mining. In: Data
  Mining, 2002. ICDM 2003. Proceedings. 2002 IEEE International Conference on.
  pp. 721--724. IEEE (2002)

\end{thebibliography}

\end{document}